\documentclass[twocolumn]{aastex631}
\usepackage{amsmath}
\usepackage{pict2e}

\makeatletter
\DeclareRobustCommand{\sector}{\mathord{\mathpalette\make@sector\relax}}
\newcommand{\make@sector}[2]{%
  \settoheight{\unitlength}{$#1x$}%
  \begin{picture}(1,1.06)
  \linethickness{.08\unitlength}
  \moveto(0.5,0)
  \lineto(0.842,1)
  \curveto(.6,1.08)(.4,1.08)(0.158,1)
  \closepath
  \strokepath
  \end{picture}%
}
\makeatother

\newcommand{\msun}{\textrm{M}_\odot}

\newcommand\pupa{RX~J0822$-$4300}
\newcommand\kes{PSR~J1852+0040}
\newcommand\tos{1E~1207.4$-$5209}

\newcommand\xmm{XMM-Newton}

\newcommand\einstein{Einstein}
\newcommand\rosat{ROSAT}

\shorttitle{Modeling the CCO in Puppis~A}
\shortauthors{Alford et al.}

\begin{document}
\title{Measuring the Non-Axially-Symmetric Surface Temperature Distribution of the Central Compact Object in Puppis A}

\correspondingauthor{Jason A. J. Alford}
\email{jason@astro.columbia.edu}

\author{J. A. J. Alford}
\affil{Columbia Astrophysics Laboratory, Columbia University, 550 West 120th Street, New York, NY 10027, USA}
\author{E. V. Gotthelf}
\affil{Columbia Astrophysics Laboratory, Columbia University, 550 West 120th Street, New York, NY 10027, USA}
\author{R. Perna}
\affil{Department of Physics and Astronomy, Stony Brook University, Stony Brook, NY 11794, USA}
\affiliation{Center for Computational Astrophysics, Flatiron Institute, 162 5th Avenue, New York, NY 10010, \
USA}
\author{J. P. Halpern}
\affil{Columbia Astrophysics Laboratory, Columbia University, 550 West 120th Street, New York, NY 10027, USA}

\begin{abstract}
The surface temperature distributions of central compact objects (CCOs) are powerful probes of their crustal magnetic field strengths and geometries.
Here we model the surface temperature  distribution of \pupa, the CCO in the Puppis~A supernova remnant (SNR), using $471$~ks of \xmm\ data.
We compute the energy-dependent pulse profiles in sixteen energy bands, fully including the general relativistic effects of gravitational redshift and light bending, to accurately model the two heated surface regions of different temperatures and areas, in addition to constraining the viewing geometry.
This results in precise measurements of the two temperatures: $kT_{\rm warm} = (1+z) \times 0.222_{-0.019}^{+0.018}$~keV and $kT_{\rm hot} = (1+z) \times 0.411\pm0.011$~keV.
For the first time, we are able to measure a deviation from a pure antipodal hot-spot geometry, with a minimum value of  $1.\!^{\circ}1 \pm 0.\!^{\circ}2$, and an expectation value of $9.\!^{\circ}35 \pm 0.\!^{\circ}17$ among the most probable geometries. 
The discovery of this asymmetry, along with the factor of $\approx2$ temperature difference  between  the  two  emitting regions, may indicate that \pupa\ was born with a strong, tangled crustal magnetic field.
\end{abstract}


\section{Introduction}

Central Compact Objects (CCOs) are a class of point-like  X-ray sources found in supernova remnants (SNRs). 
\cite{got13} confirmed the nature of CCOs as ``anti-magnetars'', i.e. neutron stars born with  weak ($10^{10-11}$~G) spin-down measured magnetic fields. 
Observationally, CCOs are characterized by steady thermal X-ray flux, a lack of emission  at other wavelengths, and the absence of a surrounding pulsar wind nebula.
We observe similar numbers of CCOs and other classes of NSs in SNRs, indicating that CCOs comprise a substantial fraction of NS births. 
Three of the eight confirmed CCOs have measured rotation periods and period derivatives.
The X-ray flux from the hot spots of these three CCOs exceeds their spin-down power, and is likely supplied by residual cooling.
The CCOs without detected X-ray pulsations may also have hot, localized surface regions that simply produce lower amplitude X-ray pulses.
See \cite{del08} and \cite{del17} for reviews of CCOs.

Puppis~A (G260.4$-$3.4) is a $\approx4600$ year old core-collapse SNR located at distance of $1.3 \pm 0.3$~kpc \citep{may20,got13,rey17}.
The oxygen enrichment of the Puppis~A ejecta indicates a very massive $\sim25$~$\msun$ progenitor star \citep{can81,hwa08}.
A point-like X-ray source near the center of the Puppis~A SNR, the CCO \pupa, was discovered by \cite{pet82} analyzing \einstein\ HRI data.
\rosat\ data later confirmed that \pupa\ is an unresolved point source, and therefore likely the compact remnant of the Puppis~A progenitor \citep{pet96}.
\cite{got09} discovered the 112~ms rotation period of \pupa, and \cite{got13} measured $\dot P=(9.28\pm0.36)\times10^{-18}$,   implying a spin-down magnetic field $B_s = 2.9 \times 10^{10}$~G.

\pupa\ has a double blackbody X-ray spectrum with temperatures of $\approx0.4$~keV and $\approx0.2$~keV, which we will call the hot and warm components.
The observed fluxes of the hot and warm components are approximately equal.
These two components, located on different regions of the star, produce X-ray pulses with strongly energy-dependent amplitudes and phases (see data in Figure \ref{fig:pp}).
At $\approx1.2$~keV, near where the hot and warm blackbodies switch dominance, the amplitude reaches a minimum and there is a $\sim180^{\circ}$ pulse phase reversal.
The combined effect of this phase reversal and the approximately equal blackbody fluxes is that the X-ray pulsations almost perfectly cancel in the integrated \xmm\ energy band.
The 112~ms rotation period of \pupa\ eluded detection until 2009, when \cite{got09} discovered its X-ray pulsations by searching in the high (1.5--4.5~keV) and low (0.5--1.0~keV) energy bands separately.
\cite{got10} found that an emission model with two perfectly antipodal hot spots, with different temperatures and areas,  successfully reproduces the observed energy-dependent pulse profiles.

In this paper, we extend the original analysis of \cite{got10} with the benefit of significantly more \xmm\ data.
We increase the number of energy bands for the pulse profile modeling from 3 to 16, and we generalize the emission model to include the possibility of a non-antipodal hot spot geometry.

\section{Data Reduction}

\begin{table}
\caption{Log of \xmm\ Observations}
\begin{tabular}{llll}
\hline 
\hline
      ObsID &        Date &  Livetime (ks) &  $Z_{1}^{2}$  \\
\hline
 0113020101 &  2001-04-15 &           15.8 &                49.3 \\
 0113020301 &  2001-11-08 &           16.1 &                51.7 \\
 0606280101-A &  2009-12-17 &           29.3 &                79.6 \\
 0606280101-B &  2009-12-17 &           26.6 &                28.4 \\
 0606280201 &  2010-04-05 &           25.3 &               101.6 \\
 0650220201 &  2010-05-02 &           13.6 &                23.7 \\
 0650220901 &  2010-10-15 &           16.4 &                46.2 \\
 0650221001 &  2010-10-15 &           16.3 &                60.7 \\
 0650221101 &  2010-10-19 &           18.6 &                42.5 \\
 0650221201 &  2010-10-25 &           17.2 &                70.2 \\
 0650221301 &  2010-11-12 &           16.5 &                25.7 \\
 0650221401 &  2010-12-20 &           19.0 &                58.2 \\
 0650221501 &  2011-04-12 &           21.0 &                68.2 \\
 0657600101 &  2011-05-18 &           25.6 &                92.3 \\
 0657600201 &  2011-11-08 &           26.1 &                51.1 \\
 0657600301 &  2012-04-10 &           24.7 &                96.8 \\
 0722640301 &  2013-10-29 &           31.9 &               118.2 \\
 0722640401 &  2013-10-31 &           29.0 &                86.2 \\
 0742040201 &  2014-10-18 &           32.1 &               106.0 \\
 0781870101 &  2016-11-08 &           50.2 &               146.9 \\
\hline
\end{tabular}

\label{table:xmm_log}
\tablecomments
{The $Z_{1}^{2}$ statistic is calculated in the 1.5--4.5 keV energy band.
The observation beginning on 2009-12-17 was split into two separate files in the \xmm\ archive, labeled here as 060628010-A and 060628010-B.  We summed all data sets to produce a single combined spectrum and a set of energy dependent pulse profiles.}
\end{table}

This study analyzes the combined EPIC-pn data from 19 \xmm\ observations of \pupa. 
A log of these observations is given in Table \ref{table:xmm_log}.
All were performed in small window mode with the thin filter.
Our analysis made use of version 16.22 of the HEASoft software, as well as version xmmsas\_20170719\_1539-16.1.0  of the \xmm\ SAS software.
Observations were reprocessed with the {\tt epchain} pipeline to apply the latest calibration products and clock corrections.
The SAS function {\tt barycen}  was used to correct photon arrival times to the solar system barycenter using the source coordinates in \cite{got13}. 
The data sets were filtered to remove time intervals contaminated with particle flares according to the recommended criteria.
Standard flag and pattern filters {\tt(PATTERN<=4 \&\& FLAG==0)} were applied.

\begin{figure*}
\centering
\includegraphics[width=.95\textwidth]{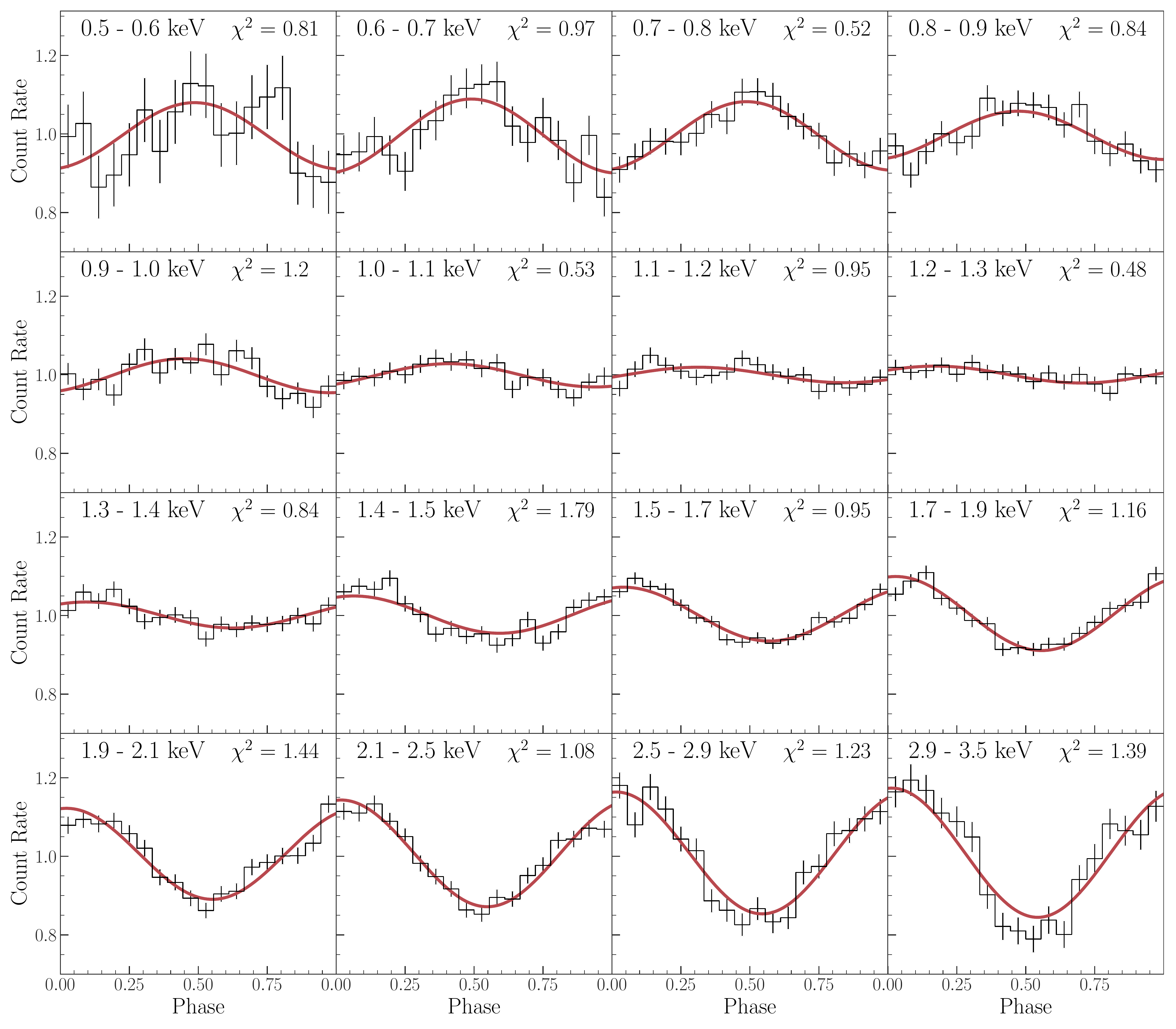}
\caption{Energy-dependent pulse profiles are plotted along with the best fit theoretical model in 16 energy bands. 
The theoretical emission model predicts the total, time-dependent X-ray flux observed in each energy band, fully including the general relativistic effects of gravitational redshift and light bending.
We have accounted for the \xmm\ detector response by folding the theoretical model through the appropriate response matrix.
Here we have subtracted background and normalized the pulse profiles to best show the energy dependence of the pulse amplitudes and phases. 
The reduced $\chi^{2}$ value for each energy band is indicated inside each subplot.
The model successfully reproduces the observed, energy-dependent pulse profiles, as well as the phase-averaged spectrum shown in Figure \ref{fig:spec}.}
\label{fig:pp}
\end{figure*}

The source photons were extracted from a circular region with a $30^{\prime\prime}$ radius, and the background spectrum was extracted from an annular region with a $32.\!^{\prime\prime}5$ inner radius and $45^{\prime\prime}$ outer radius.
We verified that larger background regions give consistent spectral results.
The size of the source region was chosen to maximize the Rayleigh statistic \citep{buc83}  in our timing analysis.
X-ray pulsations were evident in all observations, and Table \ref{table:xmm_log} shows the Rayleigh $Z_{1}^{2}$ statistic in the 1.5--4.5 keV band for each observation.
Pulse profiles from each observation were aligned by fitting the pulses in the 1.5--4.5 keV band to a sine curve and then shifting the phase of all photons so that the 1.5--4.5 keV band pulses are aligned.
The summed pulses in 16 energy bands are shown in Figure~\ref{fig:pp}.

\cite{got09} found that the spectrum of the CCO in Puppis~A shows deviations from a pure two blackbody model.
These deviations can be modeled either with the addition of an absorption line at $\approx0.45$~ keV or an emission line at $\approx0.75$~keV.
We performed a careful analysis to determine the validity of these possible spectral features.
The main concerns were that a spectral feature could be an artifact of imperfect background subtraction, or contamination of X-rays from the Puppis~A SNR.
Contamination from the supernova remnant could occur due to the details of how the pn detector operates in small window mode.
See Appendix~A for a discussion of the background characterization and subtraction method.
After concluding that the spectral feature is intrinsic to \pupa, we chose to model it as an emission line, since it yielded a marginally better fit to the phase-averaged spectrum. See also \cite{got13} for a thorough analysis and of the \pupa\ spectrum and this line feature.

\section{Emission Model}
\subsection{Defining the Viewing Geometry and Relative Positions of the Two Emitting Regions}

\begin{figure*} [h!] 
  \centering
  {\includegraphics[width=1.0\textwidth]{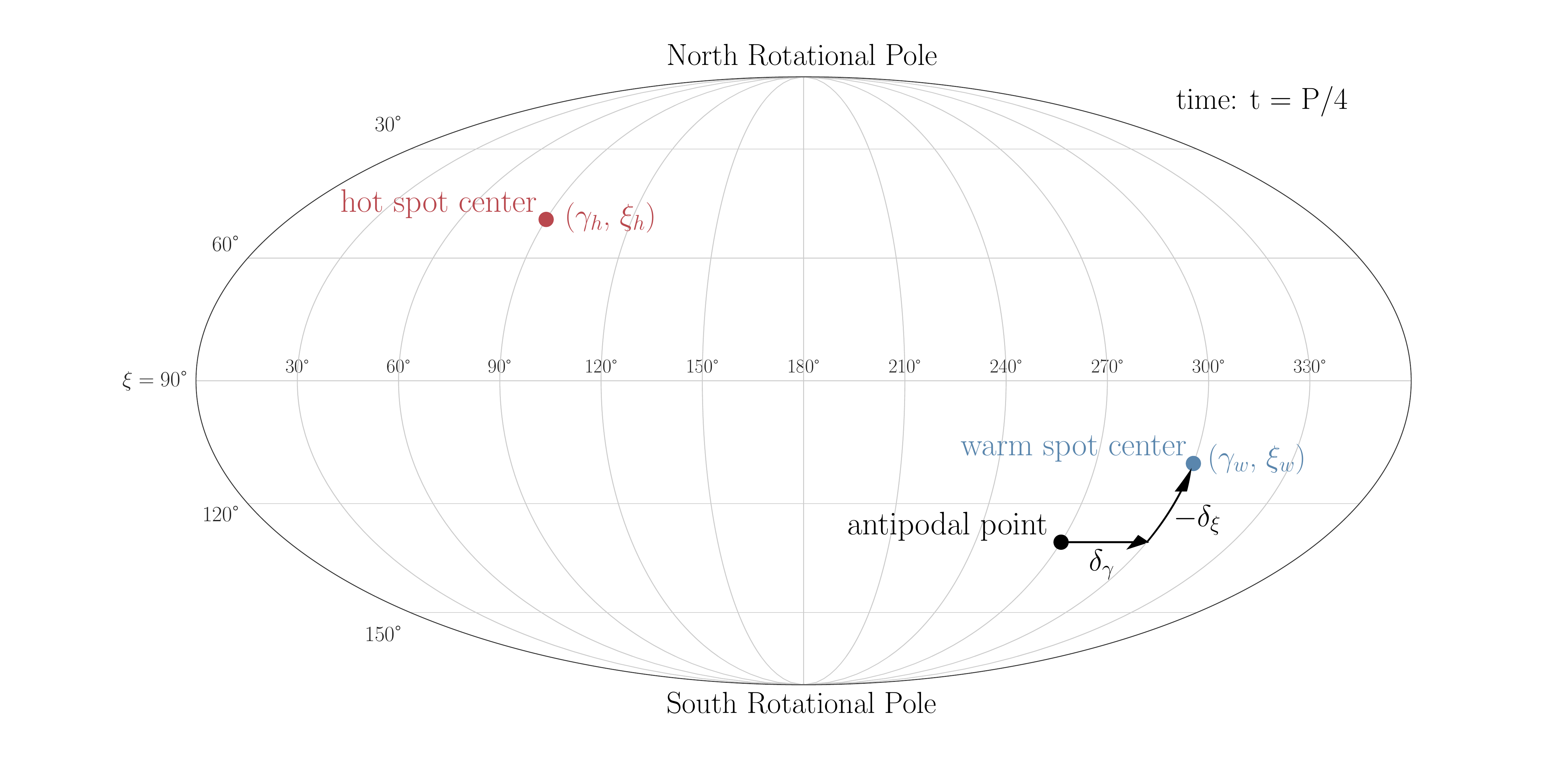}\label{fig:coord_sys}}
\caption{Mollweide projection of the coordinate system used to define the position of the warm spot relative to the hot spot. 
We define the ``North Rotational Pole'' to be the rotational pole closest to the hot spot.
The position of the hot spot also defines the line of $0^{\circ}$ longitude ($\gamma_{h}=0$ at time $t=0$).
The angle between the northern rotation axis and the hot spot is the colatitude of the hot spot, $\xi_{h}$. 
In a perfectly antipodal geometry, the warm spot would be at longitude $\gamma_{w}=\gamma_{h}+\pi$ and colatitude $\xi_{w}=\pi -\xi_{h}$, while the actual offset of the warm spot from the antipodal point is specified by $\delta_{\gamma}$ and $\delta_{\xi}$ in Equations (\ref{eq:delta_xi}) and (\ref{eq:phaselag}).
Positive/negative values of $\delta_{\xi}$ move the warm spot towards/away from the south rotational pole.
We emphasize that this coordinate system assigns colatitudes and longitudes to the positions of the hot spots \emph{at a given time t}. 
As the NS rotates, the two spots will move from left to right across lines of constant colatitude.
Longitudes and colatitudes of the ``hot'' spot and ``warm'' spot poles are indicated in parentheses.
}
\label{fig:mollweide}
\end{figure*}

Our starting point is the emission model originally described in \citet{got10}.
The observable X-ray emission from \pupa\ can be attributed entirely to two hot surface regions; the remainder of the NS surface is cool enough that it makes no detectable contribution to the X-ray spectrum observed by \xmm.
In Figure~\ref{fig:mollweide} we show the geometry of the heated regions on the surface of \pupa. 
We use a notation similar to that of \cite{got10}, labeling the angle between the rotation axis and the hot spot pole $\xi_h$, and the angle between the rotation axis and the observer's line of sight $\psi$.

The flux amplitude from the heated regions on the surface of the NS changes in time through the parameters $\alpha_h(t)$ and $\alpha_w(t)$, which are the time-dependent angles between the observer's line-of-sight and the ``hot'' spot and ``warm'' spot poles, respectively.
The phase $\gamma_{h}=0$ of the hot spot corresponds to the closest approach of the hot spot to the observer, while the phase of rotation is related to the angular rotation rate of the star $\Omega$ through $\gamma_{h}(t)=\Omega t$. 
With the spots fixed at colatitudes $\xi_h$ and
$\xi_w$, the angles $\alpha_h(t)$ and $\alpha_w(t)$ will vary with the rotation of the star as
\begin{equation}
\alpha_h(t)=\cos^{-1}[\cos\psi\cos\xi_{h}+\sin\psi\sin\xi_{h}\cos\gamma_{h}(t)]
\label{eq:alpha_h}
\end{equation}
and
\begin{equation}
\alpha_w(t)=\cos^{-1}[\cos\psi\cos\xi_{w}+\sin\psi\sin\xi_{w}\cos\gamma_{w}(t)]\,.
\label{eq:alpha_w}
\end{equation}

We use spherical caps to model the hot spots on \pupa.
We label the angular radii of the hot spot and the warm spot $\beta_{h}$ and $\beta_{w}$, respectively.
We also define two parameters, $\delta_{\xi}$ and $\delta_{\gamma}$, that specify how the  warm spot position deviates from that of the hot spot, relative to a pure antipodal geometry, in colatitude $\xi_{w}$ and longitude $\gamma_{w}$:
\begin{equation}
\xi_{w} = (\pi - \xi_{h}) + \delta_{\xi}
\label{eq:delta_xi}
\end{equation}
\begin{equation}
\gamma_{w} = \gamma_{h} + \pi + \delta_{\gamma}\,.
\label{eq:phaselag}
\end{equation}

\subsection{Intuitive Explanation of $\delta_{\xi}$, $\delta_{\gamma}$, and Degeneracies in the Emission Model}
\begin{figure*} 
\centering
\includegraphics[width=1.0\textwidth]{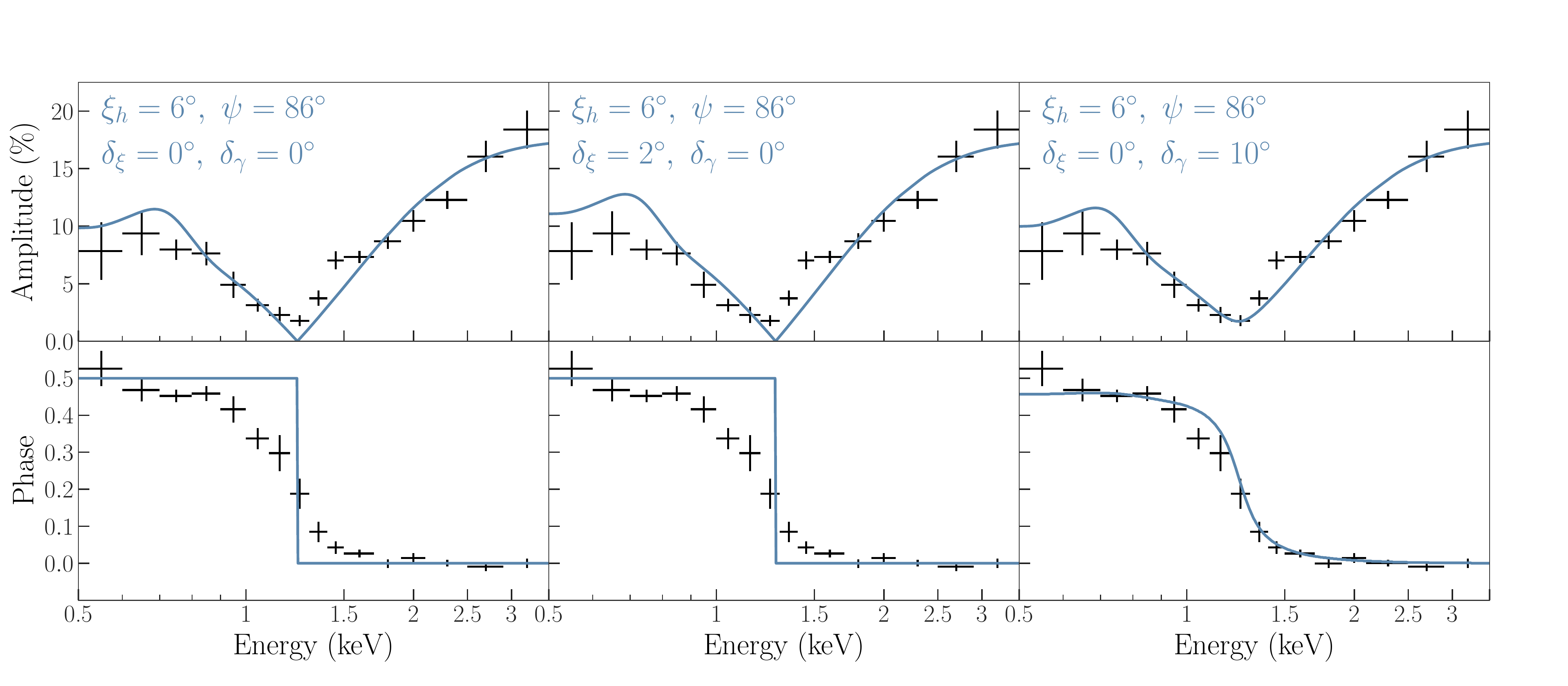}
\caption{We illustrate how the energy-dependent pulse amplitude and phase change with the model parameters $\delta_{\xi}$ and $\delta_{\gamma}$ (solid lines).
The viewing angles $\xi_{h}$ and $\psi$ are fixed at $6^{\circ}$ and  $86^{\circ}$, which are the values \cite{got10} calculated for the perfectly antipodal model.
The phase plot must be a step function if $\delta_{\gamma} = 0$. The smoother shift
in phase of the data points demonstrates that the pulse profiles of \pupa\ are inconsistent with an antipodal geometry.
The $\delta_{\gamma} \neq 0^{\circ}$ condition also correctly predicts the nonzero pulse amplitude that we observe in all energy bands.
Fitting only the amplitude and phase to the observed data is not sufficient for accurate modeling, which requires fitting the exact pulse shape in each energy band as shown in Figure~\ref{fig:pp}.
The theoretical amplitude and phase curves calculated from the best fit parameters are shown in the lower panels of Figure \ref{fig:spec}.
}
\label{fig:amplitude_phase}
\end{figure*}

Adjusting the values of $\delta_{\xi}$ and $\delta_{\gamma}$, the warm spot can be placed anywhere on the NS relative to the hot spot.
In the special case of an antipodal geometry, $\delta_{\xi}=0$ and $\delta_{\gamma}=0$.
Figure \ref{fig:amplitude_phase} shows qualitatively how the energy-dependent pulse amplitude and phase are affected by the parameters $\delta_{\xi}$ and $\delta_{\gamma}$.
In our coordinate system, the amplitude and phase of the X-ray pulses from the hot spot are independent of both $\delta_{\xi}$ and  $\delta_{\gamma}$.
Also, Equation~\ref{eq:alpha_w} implies that $\delta_{\gamma}$ determines the phase of the X-ray pulses from the warm spot, while $\delta_{\xi}$ affects only the amplitude of the X-ray pulses from the warm spot.
If the fluxes from the two spots are comparable and $\delta_{\gamma}=0$, then the observed X-ray pulse phase as a function of energy will be a step function.  
But, as we shall show, $\delta_{\gamma} \neq 0$ is required to match the observations. 

As discussed in \cite{got10}, for a perfectly antipodal geometry, and with $\psi \leq 90^{\circ}$, there is a degeneracy in the interchange of the viewing angles $\xi_{h}$ and $\psi$. 
However, when we generalize the model to allow for non-antipodal spots and extend the range of $\psi$ up to $180^{\circ}$, this degeneracy is broken, and another degeneracy emerges involving the viewing angles $\xi_{h}$, $\psi$, and the dipole offset angle $\delta_{\xi}$. 
The following is an intuitive explanation of this degeneracy: There is a curve in the $\psi$, $\xi_{h}$  parameter space where the hot-spot pulse amplitude is constant and equal to what is observed.
At all points on this curve the position of the warm spot can be adjusted so that its pulse amplitude matches what is observed.
Both spot sizes vary independently along this curve, so that the \emph{observed} flux from each spot is constant.
The angle $\delta_\gamma$ doesn't participate in this degeneracy because it only affects the relative phases of the X-ray pulses, and not the amplitude of the individual pulses.

\subsection{Light Bending and Gravitational Redshift}
We adopt a spherical coordinate system where the colatitude angle $\theta$ is measured with respect to the observer's line of sight.
General relativity predicts that a photon traveling from a colatitude $\theta$ on the NS surface will reach an observer only if it was emitted at an angle $\delta$ measured from the surface normal.
The light bending angle $\theta$ as a function of the emission angle $\delta$ is given by the following elliptic integral \citep{pec83}:
\begin{equation}
\theta(\delta)=\int_0^{R_s/2R}x\;du\left/\sqrt{\left(1-\frac{R_s}{R}\right)
\left(\frac{R_s}{2R}\right)^2-(1-2u)u^2 x^2}\right.,
\label{eq:theta_exact}
\end{equation}
where $x\equiv \sin\delta$, $R/R_s$ is the NS radius in units of the Schwarzschild radius $R_s=2GM/c^2$.
To improve the efficiency of this calculation we use an approximation presented in \cite{bel02}:
\begin{equation}
\label{eq:cos}
    1-\cos\delta=(1-\cos\theta)\left(1-\frac{R_s}{R}\right).
\end{equation}
For a $1.4~\msun$ NS, with any reasonable NS radius, the error introduced by this approximation is $\lesssim 1\%$, smaller than the statistical uncertainties in the data.

The  observed spectrum as a function of rotational phase is computed by integrating over the visible area of the hot spots according to 
the formula given in \cite{bel02}:
\begin{equation}
\label{eq:flux_dF}
  dF=\left(1-\frac{R_s}{R}\right)^2I_0(\delta)\cos\delta\frac{dS}{D^2}\,.
\end{equation}
Here $dF$ is the flux from a surface element $dS$, $I_0(\delta)$ is the intensity in the NS rest frame as a function of the emission angle $\delta$, and $D$ is the distance to the NS.

\subsection{Flux Integration}

The observed pulse profiles of \pupa\ are consistent with a sinusoid, indicating that the intensity is consistent with being isotropic (i.e. $I_0(\delta)$ is actually independent of $\delta$). 
In this special case of isotropic intensity, the integration of Equation \ref{eq:flux_dF} is a simple calculation of the average value of $\cos\delta$ on the NS hot spot (and the same for the warm spot), 

\begin{multline} \label{eq:avg_cos_delta}
<\cos \delta>  = \int_{\alpha_{h}-\beta_{h}}^{{\rm min}(\alpha_{h}+\beta_{h}, \theta_{\rm{max}} ) }  d\theta  \cos\delta \\
\times \cos^{-1}\left(\frac{\cos\beta_h-\cos\alpha_h\cos\theta}{\sin\alpha_h\sin\theta}\right),
\end{multline}
where $\theta_{\rm{max}}$ is the maximum NS colatitude visible to an observer.
Equation \ref{eq:cos} implies that $\theta_{\rm{max}} = 121.\!^{\circ}7$ for a 12~km radius, 1.4~$\msun$ NS.

When calculating the two dimensional integral of $dF$ over the visible area of a hot spot, there is a trade-off between computational accuracy and speed. It is preferable to calculate only the one dimensional integral above, and use an exact analytic formula for the visible hot spot area, if such a formula exists for the shape of the hot spot.

Here we calculate the visible surface areas of the spherical caps that we use to model the hot spots on \pupa.
The area of a spherical cap with angular radius $\beta$ on a NS with radius $R_{NS}$ is $2\pi(1-\cos\beta) R_{NS}^{2}$, so this is the value of $\int dS$ when the entire cap is visible.
When the spherical cap is only partially visible, we use the exact analytic formula for the area of intersection $I(\theta, r_1, r_2)$ of two spherical caps with angular radii r$_{1}$ and $r_{2}$, separated by an angle $\theta$ on a unit sphere (see derivation in Appendix~B):
\begin{multline} 
I(\theta, r_1, r_2) = 2 \pi \\
-2 \cos^{-1}\bigg(\frac{\mathrm{cos}\ \theta}
   { \mathrm{sin}\ r_1\ \mathrm{sin}\ r_2} -
   \frac{1}{\mathrm{tan}\ r_1\ \mathrm{tan}\ r_2} \bigg) \\
\ - \ 2 \cos^{-1}\bigg(\frac{\mathrm{cos}\ r_2}
   { \mathrm{sin}\ \theta\ \mathrm{sin}\ r_1} -
   \frac{1}{\mathrm{tan}\ \theta\ \mathrm{tan}\ r_1} \bigg) \mathrm{cos}\ r_1  \\
 \  - \ \ 2 \cos^{-1}\bigg(\frac{\mathrm{cos}\ r_1}
   { \mathrm{sin}\ \theta\ \mathrm{sin}\ r_2} -
   \frac{1}{\mathrm{tan}\ \theta\ \mathrm{tan}\ r_2} \bigg)
   \mathrm{cos}\ r_2
\label{eq:cap_intersection}
\end{multline}
So, the general formula for the visible area $\int dS$ of a spherical cap with angular radius $\beta$ at colatitude $\alpha$ is
\begin{equation}
\int dS = \left\{\begin{array}{ll} 
2\pi(1-\cos \beta) R_{NS}^{2} & \alpha + \beta < \theta_{\rm max} \\
I(\pi-\alpha, \beta, \theta_{\rm max})R_{NS}^{2} & \alpha - \beta < \theta_{\rm max} < \alpha + \beta \\
0 & \alpha - \beta > \theta_{\rm max}
\end{array}\right. 
\label{eq:exact_cap_area}
\end{equation}

In the following section we will fit our emission model to the \xmm\ data and constrain the viewing geometry of \pupa.
We will accurately measure the temperatures of the two hot spots, calculate the minimum and maximum possible values of the deviation from a perfectly antipodal configuration, and calculate the corresponding values of the hot and warm spot sizes.

\section{Energy-Dependent Pulse Profile Modeling}
\subsection{Accurately Measuring the Temperatures of the Two Hot Spots}

Our modeling procedure first constrains the temperatures of the two hot spots, and then constrains the values of the geometric angles $\psi$, $\xi_{h}$, $\delta_{\xi}$, and $\delta_{\gamma}$.
The amplitudes and phases of energy-dependent pulse profiles are strongly dependent on the temperatures of the two hot spots.
This is especially true around the 1.2~keV region where the pulses from the two spots almost perfectly cancel out.
The extreme sensitivity of the pulse profiles to small changes in the individual hot spot temperatures implies that very accurate measurements of the two hot spot temperatures will result from our modeling.

We follow a procedure that uses information from the pulse profiles to measure the spot temperatures much more accurately than what can be achieved through spectral modeling alone.
We search through different temperatures of the warm and hot spots by holding the warm spot temperature fixed and allowing the rest of the model parameters to vary to fit the phase-averaged spectrum.
We repeated this for a range of values of the redshifted warm spot temperature, varying the temperature from 0.18 to 0.28~keV in steps of 0.001~keV, and calculating the other model parameters that best fit the phase-averaged spectrum for each fixed warm spot temperature.
So, after this calculation, for each value of $kT_{\rm warm}$, we have the values of the $kT_{\rm hot}$, $L_{\rm warm}$, $L_{\rm hot}$, $E_{\rm line}$, $L_{\rm line}$, and $N_{\rm H}$ that best fit the phase-averaged spectrum.

For each set of warm spot/hot spot temperature pairs consistent with the phase-averaged spectra, we fold the absorbed spectrum of each spectral component through the \xmm\ detector response matrix and compute the phase-averaged flux in each of the 16 energy bands.
We divide each energy band into 18 phase bins.
Since the goal of this procedure is only to accurately measure the spot temperatures, and not small deviations of their spectra from perfect blackbodies, we employ a flux ``renormalization'' method that adjusts the flux in each energy band to match what is observed, since we are trying to extract information about the temperatures from the pulse shape and are not concerned about small statistical fluctuations in the normalization. 
In all energy bands the flux adjustment is smaller than the $2\%$ \xmm\ pn detector systematic uncertainties\footnote{\label{fn:note1}\url{https://xmmweb.esac.esa.int/docs/documents/CAL-TN-0018.pdf}}.

\begin{figure*}
 \centering
\includegraphics[width=0.95\linewidth]{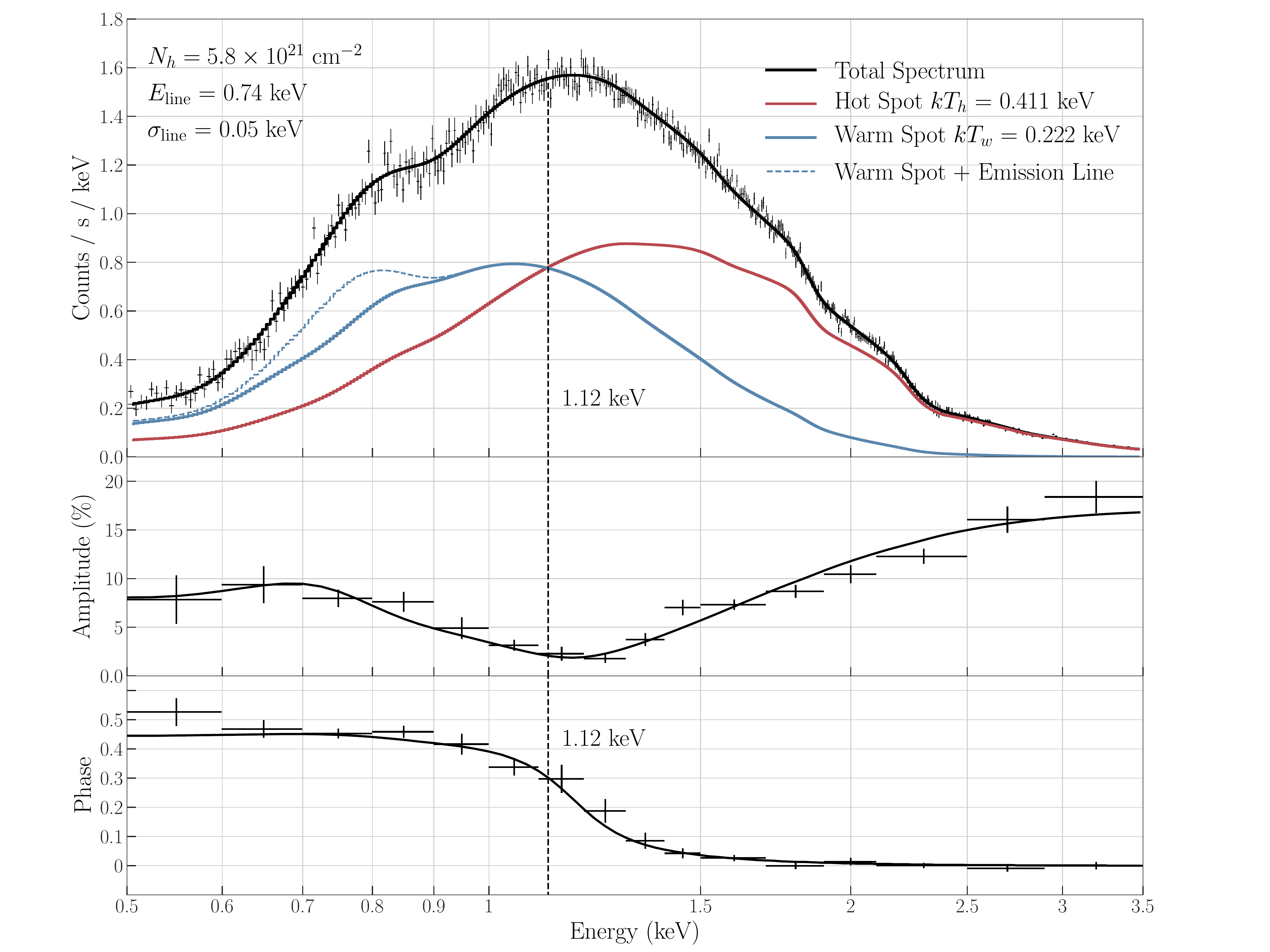}
\caption{\label{fig:spec}
Top panel: The phase-averaged spectrum of \pupa, plotted along with the best fit model. 
The individual model components from the hot spot, warm spot, and emission-line feature are shown. 
The temperatures are calculated to high precision by combining the spectral data with the additional constraints required to match the energy-dependent pulse profiles.
Table \ref{tbl:obs_params} lists the best fit values of phase-averaged spectral model parameters and their uncertainties.
Middle panel: The observed pulse amplitudes are plotted along with the amplitudes predicted by our model.
The observed fluxes of the hot and warm components are equal at 1.12~keV, and the pulse amplitude is minimized at a slightly higher energy: 1.17~keV.
This is because the intrinsic amplitude of the warm component ($20\%$) is larger than the hot component amplitude ($17\%$).
Bottom panel: The observed pulse phases are plotted along with the phases predicted by our model.
The longitude of the warm spot on the NS, $\gamma_{w}$, differs from the hot spot longitude $\gamma_{h}$ according to Equation \ref{eq:phaselag}: $\gamma_{w} = \gamma_{h} + \pi + \delta_{\gamma}$.
We find that $\delta_{\gamma} = 11.\!^{\circ}7$, which accurately reproduces the energy-dependent phases plotted in the lower panel.
}
\end{figure*}

For each possible spectral model, parameterized by $kT_{\rm warm}$, we search for the amplitudes and phases that best fit the observed, energy-dependent pulse profiles.
We perform this search using simple sine waves to model the modulation of the hot and warm components.
For most geometries, this is actually not an approximation beyond the ``cosine'' light bending approximation (Equation \ref{eq:cos}) that we are already using. 
In the case where the intensity $I_0(\delta)$ is isotropic, and both hot spots are always entirely visible to an observer, Equations \ref{eq:alpha_h} and \ref{eq:alpha_w} combined with Equation \ref{eq:flux_dF} imply that the pulses from both spots are \emph{exactly} simple sine waves.
We will find that the most probable emission geometries for \pupa\ correspond to this most simple case.
However, we also have to consider the case where a hot spot is only partially visible to an observer during a rotation of the NS.
Then, the flux contributions from the hot spot surface elements $dS$ that are always visible will still be simple sine waves.
The the flux contributions from the hot spot surface elements $dS$ that are not always visible will be ``truncated'' sine waves.
In the regions of the parameter space that produce the observed low amplitude X-ray pulses, these truncated sine wave contributions are a negligible fraction of the observed flux, since they correspond to surface elements $dS$ where $<\cos \delta>$ is smallest.
It turns out that the fractional error from an exact sine wave is very small $\sim 10^{-4}$.
So, even in these special cases, the simple sine wave pulse model is appropriate for our purpose here of determining the intrinsic pulse amplitudes and redshifted temperatures.
We find that the hot and warm spot pulse amplitudes are $16.9 \pm 0.8\%$ and $20.0 \pm 1.4\%$, respectively.
We identify the pair of temperatures ($kT_{\rm warm}$, $kT_{\rm hot}$) that are best able to reproduce the energy-dependent pulse profiles as the correct, redshifted temperatures.
We find that  $kT_{\rm warm} = 0.222_{-0.019}^{+0.018}$~keV and $kT_{\rm hot} = 0.411\pm0.011$~keV.
We emphasize that these are the values of the redshifted temperatures that best fit \emph{both the pulse profiles and the phase-averaged spectra}, and are not the same as the values obtained by only fitting the phase-averaged spectra.
The parameters of this phase-averaged spectral model are listed in Table \ref{tbl:obs_params}, and the phase-averaged spectrum is shown in Figure~\ref{fig:spec}.
 
\subsection{Computing the "Beta Maps"}
Having the correct hot spot temperatures (as seen by an observer at infinity), as well as the intrinsic pulsation amplitudes and relative phase of the two spectral components, we can then determine the corresponding possible values of the viewing angles $\xi_{h}$ and $\psi$, and also the position of the warm spot relative to the hot spot parameterized by the angles $\delta_{\xi}$ and $\delta_{\gamma}$.
In Section \ref{section:param_search}, we will compute the time-dependent spectra from our general relativistic emission model over the four-dimensional parameter space of all plausible combinations of the physical parameters $\xi$, $\psi$, $\delta_{\xi}$, and $\delta_{\gamma}$.
In order to compute these spectra, it is necessary to know the values of the spot sizes $\beta_{w}$ and $\beta_{h}$ for all of the possible values of $\xi$, $\psi$, $\delta_{\xi}$, and $\delta_{\gamma}$.
Since $\delta_{\gamma}$ only affects the relative phases of the pulses from the hot and warm spots, the sizes of both spots are independent of $\delta_{\gamma}$.  The parameters
$\beta_{w}$ and $\beta_{h}$ are also functions of the mass, radius and distance to \pupa.
We assume $M_{NS} = 1.4~\msun$, $R_{NS} = 12$~km and $D = 1.3$~kpc.
While $\beta_{w}$ and $\beta_{h}$ do depend on these parameters, it will turn out that the geometric angles we are trying to measure are not very sensitive to the assumed distance and NS radius.

\begin{figure*}   
\includegraphics[width=1.\linewidth]{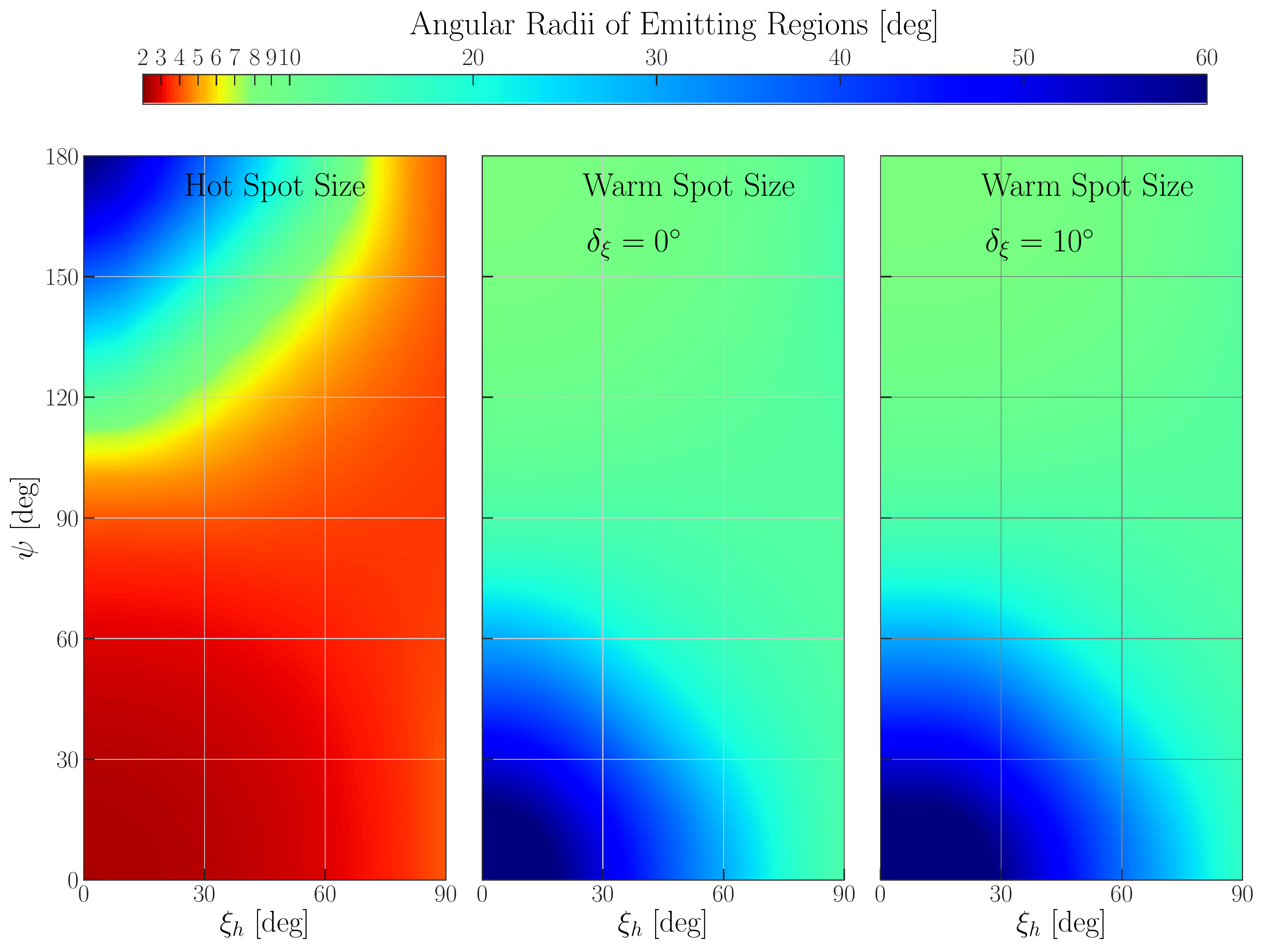}
 \caption{Sizes of the emitting regions as a function of all possible viewing angles, $\psi$ and $\xi_{h}$, and the relative positions of the heated regions on the NS surface, parameterized by the angles $\delta_{\xi}$ and $\delta_{\gamma}$. 
In our coordinate system the hot-spot size is independent of $\delta_{\xi}$ and $\delta_{\gamma}$.
The warm-spot size is a function of $\delta_{\xi}$, but is independent of $\delta_{\gamma}$. 
We show two representative plots of the warm-spot sizes.
Left: The hot spot angular radius as a function of the viewing angles $\psi$ and $\xi_{h}$.
Center: Warm spot angular radius with $\delta_{\xi}$ = 0.
Right: Warm spot angular radius  with $\delta_{\xi} = 10^{\circ}$.
The spot sizes shown here are smaller than the corresponding  sizes in \cite{got10} because we are using the updated, closer distance estimate of 1.3~kpc \citep{rey17}. }
\label{fig:beta_maps}
\end{figure*}

Figure \ref{fig:beta_maps} shows some examples of these ``beta maps'', i.e., maps of the sizes of the hot and warm spot as a function of the viewing angles $\xi_{h}$ and $\psi$. 
There is only one beta map for the hot spot because its position relative to the observer only depends on $\xi_{h}$ and $\psi$.
For the warm spot we construct beta maps for all values of the parameter $\delta_{\xi}$.

\subsection{Searching the Parameter Space}
\label{section:param_search}
For each pair of the parameters ($\xi_{h},\psi$) and ($\delta_{\xi},\delta_{\gamma}$), the ``beta maps'' give the corresponding hot-spot sizes, and we can now compute the values of the two pulsation amplitudes and the relative phase from our emission model.
Using this mapping between the physical parameters of the NS geometry, and the two pulsation amplitudes and the relative phase of the two spectral components, we calculate the pulse profiles corresponding to each set of physical parameters.
We then record the $\chi^2$ values obtained from comparing each set of theoretical energy-dependent pulse profiles to the data.
The best fit of the theoretical pulse profiles to the data yields a reduced $\chi^2_{\nu}=1.01$ for $\nu=284$ degrees-of-freedom.
Figure \ref{fig:pp} shows the theoretical model superposed on the energy-dependent pulse profiles.

\begin{figure*} [h!]
{\includegraphics[width=1.0\textwidth]{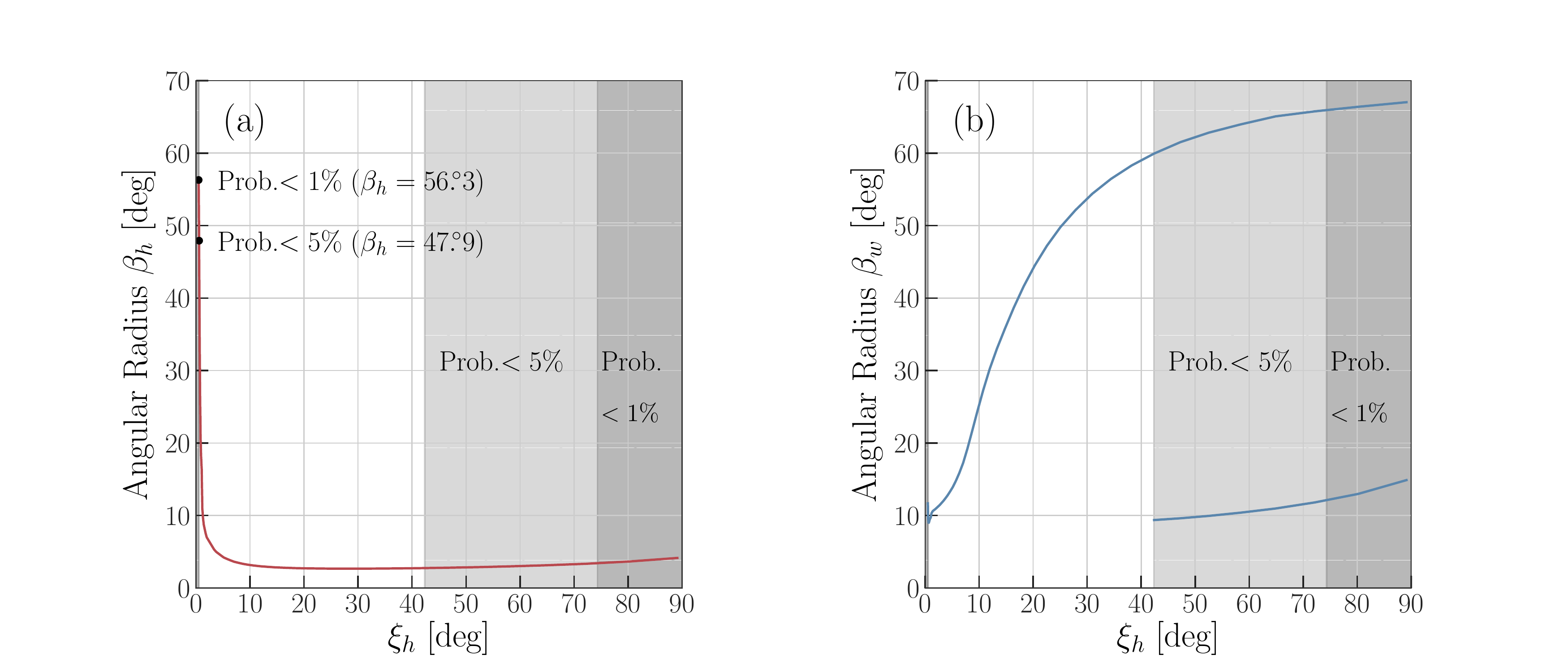}}
\caption{Left: Hot spot size as a function of the hot spot colatitude $\xi_{h}$.
The ranges of $\xi_{h}$ that can be ruled out at $95\%$ and $99\%$ confidence are shaded gray.
These ranges are difficult to see at small values of $\xi_{h}$, so the corresponding hot spot sizes $\beta_{h}$ are indicated.
Right: Warm spot sizes parameterized by the hot spot colatitude $\xi_{h}$.
The ranges of $\xi_{h}$ that can be ruled out at $95\%$ and $99\%$ confidence are shaded gray.
At values of $\xi_{h} > 42.\!^{\circ}4$ there are two possible locations and therefore sizes of the warm spot.
}
\label{fig:betas}
\end{figure*}

We derive constraints on $\xi_{h}$, $\psi$, $\delta_{\xi}$ and $\delta_{\gamma}$ by calculating $\Delta\chi^2$ above the best fit value.
Because $\delta_{\gamma}$ only affects the relative phases of the X-ray pulses from the warm and hot spots, its value can be unambiguously measured.
We find that $\delta_{\gamma}= 11.\!^{\circ}7^{+2.\!^{\circ}6}_{-2.\!^{\circ}5}$.
Due to the added degeneracy of the offset dipole emission model, the range of allowed values of $\xi_{h}$ and $\psi$ is substantially larger than in the perfectly antipodal model considered by \cite{got10}.
In fact, for all hot spot positions  $0.\!^{\circ}5 < \xi_{h} < 89.\!^{\circ}5$, there is exactly one viewing angle $\psi$ that gives the correct hot spot pulsation amplitude.
In turn, for each of these values of $\psi$, there is \emph{at least} one value of $\delta_{\xi}$ that gives the correct warm spot pulsation amplitude.
While this is a large range of model parameters that can reproduce the observed X-ray data, in the next section we will show that all but a very limited range of these parameters are very improbable.
Figure \ref{fig:degeneracy}a shows the full range of allowed combinations of $\xi_{h}$ and $\psi$, along with the one-sigma uncertainties in the angle $\psi$.
Figure \ref{fig:degeneracy}b shows the values of the dipole offset angle $\delta_{\xi}$ and its one-sigma uncertainties, parameterized by the hot spot inclination angle $\xi_{h}$.
Figure \ref{fig:betas} shows the sizes the hot and warm spots as a function of the hot spot colatitude $\xi_{h}$.

When $\xi_{h} > 42.\!^{\circ}4$ (or equivalently when $\psi < 18.\!^{\circ}1$), there are two allowed values of $\delta_{\xi}$ for each ($\xi_{h}, \psi)$ pair.
These two curves are shown in Figure~\ref{fig:degeneracy}b.
In the upper curve, when $\xi_{h} > 42.\!^{\circ}4$, the center of the warm spot is always $< 3^{\circ}$ from the south rotational pole.
In the lower curve, when $\xi_{h} > 42.\!^{\circ}4$, the center of the warm spot approaches the rotational equator as $\xi_{h}$ approaches $90^{\circ}$.

\begin{figure*}[h]
 \centering
\includegraphics[width=0.5\linewidth]{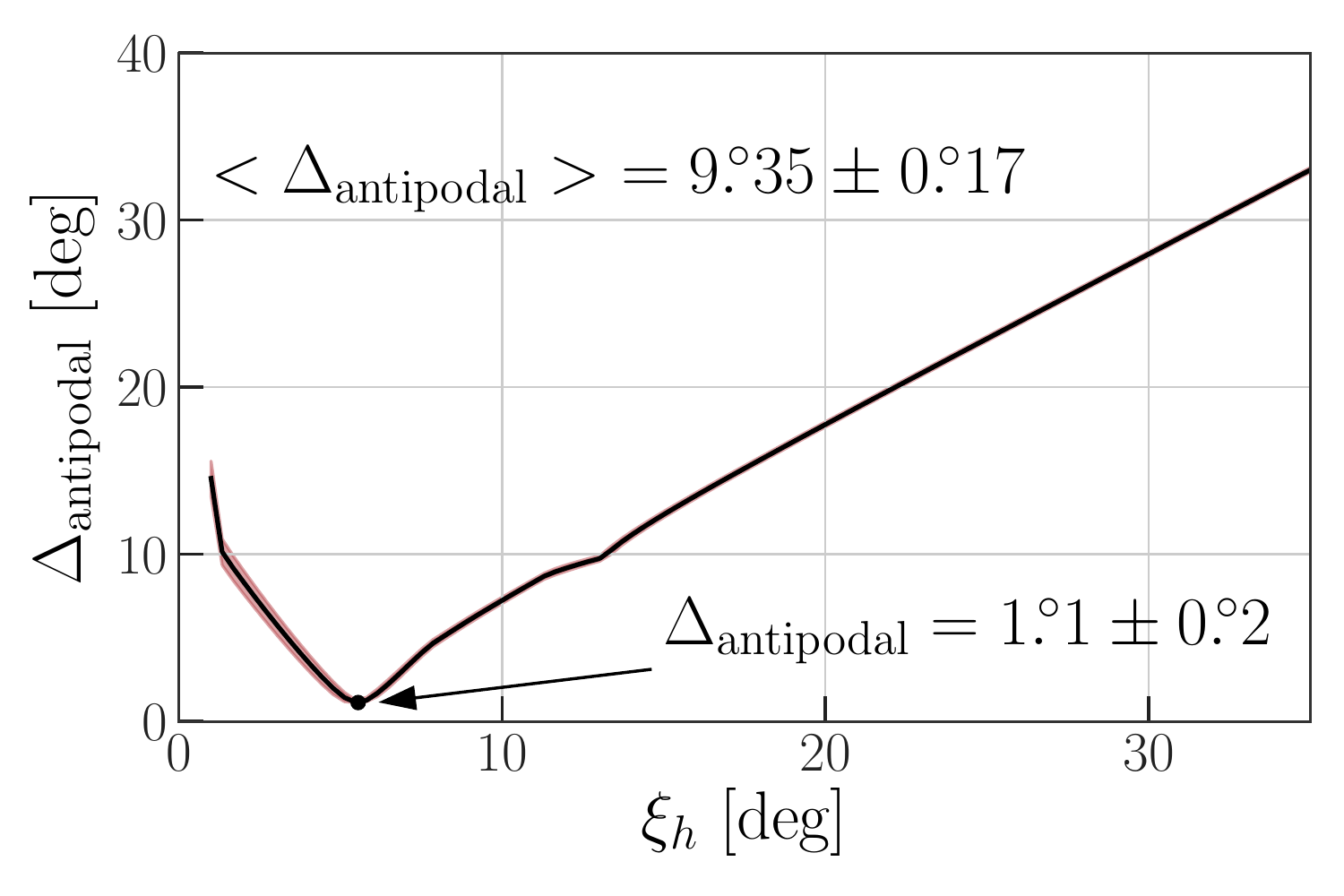}
\caption{\label{fig:antipodal_offset}
Angular distance $\Delta_{\rm antipodal}$ between the center of the warm spot and the antipodal point.
We plot $\Delta_{\rm antipodal}$ for the most probable values of $\xi_{h}$, $1^{\circ} \leq  \xi_{h} \leq 35^{\circ}$, computed in Section \ref{subsection:fine_tuning}.
The one-sigma uncertainty in $\Delta_{\rm antipodal}$ is indicated by the red shaded region, which is sometimes smaller than the width of the line.
At ($\xi_{h}$, $\psi$) = ($5.\!^{\circ}5$, $87.\!^{\circ}9$), we find the minimum value: $\Delta_{\rm antipodal} = 1.\!^{\circ}1 \pm 0.\!^{\circ}2$.
Averaging over the most probable geometries, we calculate the expectation value $<\Delta_{\rm antipodal}> = 9.\!^{\circ}35 \pm 0.\!^{\circ}17$.}
\end{figure*}

We note here three special configurations of the two emitting spots.
We have labeled these special configurations in Figure \ref{fig:degeneracy}a.
First, there is a configuration where the two spots have equal sizes.
This corresponds to the angles ($\xi_{h}$, $\psi$, $\delta_{\xi}$) = ($1.\!^{\circ}2$, $115.\!^{\circ}9$, $-10.\!^{\circ}9$).
Second, there are two degenerate configurations where $\delta_{\xi}=0$.
These two configurations correspond to the angles ($\xi_{h}$, $\psi$) = ($5.\!^{\circ}5$, $87.\!^{\circ}9$) and ($\xi_{h}$, $\psi$) = ($87.\!^{\circ}9$, $5.\!^{\circ}5$).
As expected, these two configurations intersect the one-sigma confidence intervals of the perfectly antipodal solutions calculated in \cite{got10}.
The first of these configurations, ($\xi_{h}$, $\psi$) = ($5.\!^{\circ}5$, $87.\!^{\circ}9$), corresponds to the minimum possible offset between the center of the warm spot and antipodal point.
This distance, $\Delta_{\rm antipodal}$, is measured along the great circle connecting the the center of the warm spot and the antipodal point.
At ($\xi_{h}$, $\psi$) = ($5.\!^{\circ}5$, $87.\!^{\circ}9$), we calculate that $\Delta_{\rm antipodal} = 1.\!^{\circ}1 \pm 0.\!^{\circ}2$.
If we consider the most probable geometries (discussed in Sections \ref{subsection:Most_probable_geometries} and \ref{subsection:fine_tuning}), and calculate the expectation value of this distance, we find that: $<\Delta_{\rm antipodal}> = 9.\!^{\circ}35 \pm 0.\!^{\circ}17$.
Figure \ref{fig:antipodal_offset} shows $\Delta_{\rm antipodal}$ as a function of $\xi_{h}$, for the most probable values of $\xi_{h}$.

\subsection{Identifying the Most Probable Geometries}
\label{subsection:Most_probable_geometries}
\begin{figure*} [h!] 
  \centering
  {\includegraphics[width=1.0\textwidth]{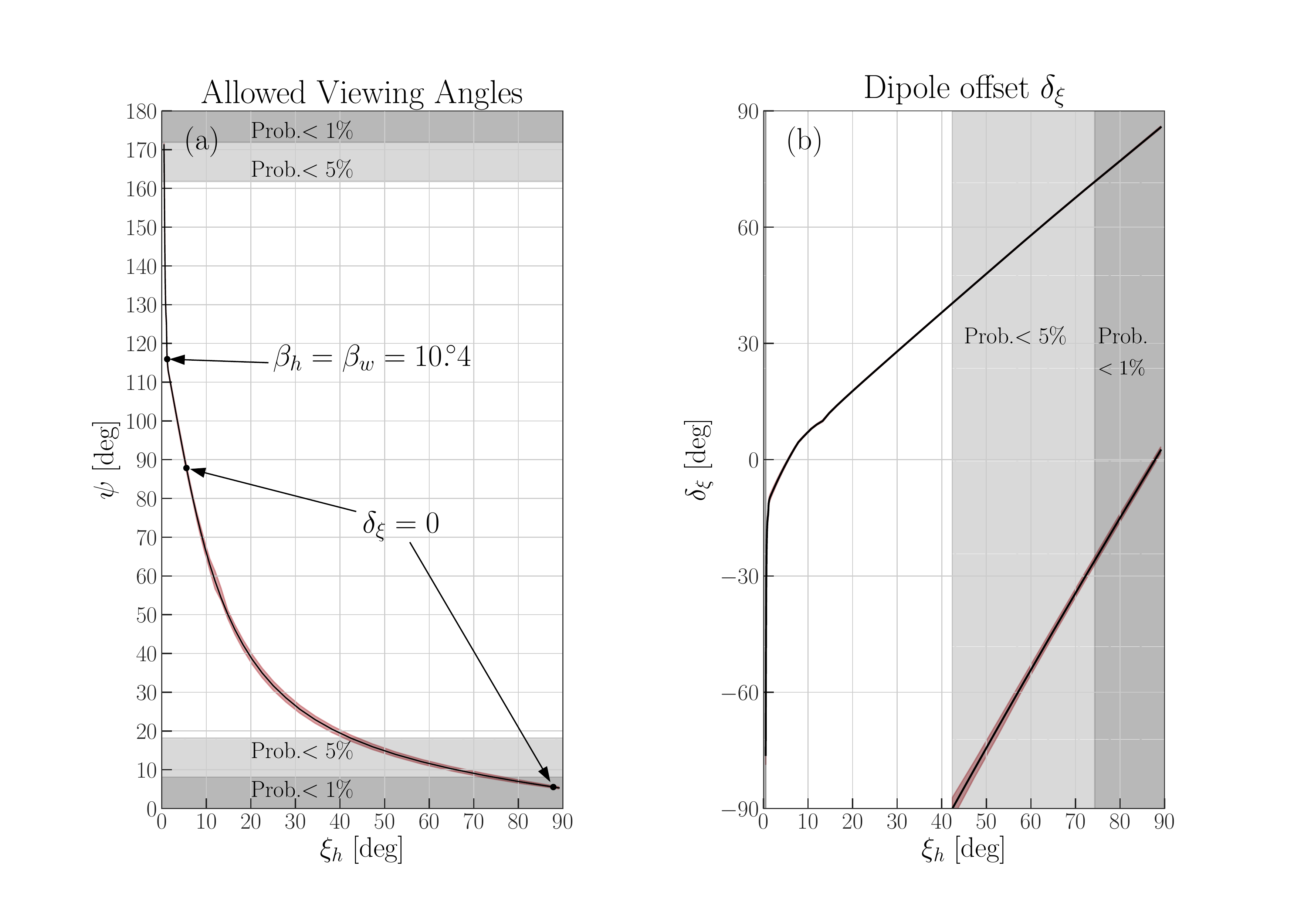}}
\caption{Constraints on the viewing geometry and dipole offset parameters of \pupa.
These angles are conveniently parameterized by $\xi_{h}$, the angle between the center of the hot spot and the rotation axis.
The one-sigma uncertainties in the angles $\psi$ and $\delta_{\xi}$  are indicated by the red shaded regions, which are sometimes smaller than the width of the line.
Left: The values of $\psi$, the angle between observer's line of sight and the rotation axis, as a function of $\xi_{h}$.
The ranges of $\psi$ that can be ruled out at $95\%$ and $99\%$ confidence are shaded gray.
The points where the hot and warm spots are the same size, and where the dipole offset angle is smallest, are indicated.
Right: The values of the dipole offset angle $\delta_{\xi}$, parameterized by the angle $\xi_{h}$. 
The ranges of $\xi_{h}$ that can be ruled out at $95\%$ and $99\%$ confidence are shaded gray.
At values of $\xi_{h} > 42.\!^{\circ}4$ there are two possible locations of the warm spot.
In the upper curve, when $\xi_{h} > 42.\!^{\circ}4$, the center of the warm spot is always $< 3^{\circ}$ from the south rotational pole.
In the lower curve, when $\xi_{h} > 42.\!^{\circ}4$, the center of the warm spot approaches the rotational equator as $\xi_{h}$ approaches $90^{\circ}$. 
}
\label{fig:degeneracy}
\end{figure*}

While a range of values of $\xi_{h}$ and $\psi$ are consistent with the data, large regions of this parameter space are highly improbable.
First we consider $\psi$, the angle between the rotation axis and the observer's line of sight, which should be a sinusoidally distributed random variable.
The reason is that this is the distribution of angles one would get by sampling the angles between two vectors pointing in random directions in 3D space.
In this case the two vectors are the NS spin axis and our line of sight to the NS.
This means that there is a $99 \%$ probability that $8.\!^{\circ}11 < \psi < 171.\!^{\circ}89$, and a $95 \%$ probability that $ 18.\!^{\circ}19 < \psi < 161.\!^{\circ}81$.
Figure \ref{fig:degeneracy} shows the $<1 \%$ and $<5 \%$ probability regions of the parameter spaces shaded in gray.
In particular, there is a $<5 \%$ probability that the true geometry of \pupa\ corresponds to a point on one of the two $\delta_{\xi}$ curves where $\xi_{h} > 42.\!^{\circ}4$.

As $\xi_{h}$ approaches $0^{\circ}$, $\psi$ approaches $180^{\circ}$, and the size of the hot spot $\beta_{h}$ increases rapidly for values of $\xi_{h} \lesssim 1^{\circ}$.
The reason for this trend is that at these large values of $\psi$, the hot spot is mostly invisible to an observer, and the observer is seeing only a small edge of the hot region.
We consider these geometries highly improbable because all of these scenarios involve ``fine tuning'' the size of the hot spot so that just enough of it is visible to an observer to roughly match the flux of the warm spot.
See Section \ref{subsection:fine_tuning} for a discussion of this ``fine tuning''.

\begin{table}[htb]
\caption{Observed spectral parameters of \pupa.}
\begin{center}
    \begin{tabular}{ l | l  }
        \toprule
        Parameter & Value  \\
        \toprule
        $N_{\rm H}$ & $0.58_{-0.02}^{+0.01}$ $\times 10^{22}$ cm$^{-2}$  \\
        $kT_{\rm warm}$ & $0.222_{-0.019}^{+0.018}$ keV  \\
        $L_{\rm warm}$ & $9.5_{-0.3}^{+0.4} \times 10^{32}$ $D_{1.3}^{2}$ erg s$^{-1}$  \\
        $kT_{\rm hot}$ & $0.411\pm0.011$  keV  \\
        $L_{\rm hot}$ & $(1.01\pm0.01) \times 10^{33}$ $D_{1.3}^{2}$ erg s$^{-1}$  \\
        $E_{\rm line}$ & $0.74\pm0.01$ keV  \\
        $\sigma_{\rm line}$ & 0.05 keV (fixed)  \\
        $L_{\rm line}$ & $5.8_{-1.0}^{+1.2}$ $\times 10^{31}$ erg s$^{-1}$  \\
        \hline
    \end{tabular}
    \end{center}
\tablecomments{
$D_{1.3}$ is the distance to \pupa \ in units of 1.3~kpc. 
The temperatures of the two blackbodies and the central energy of the emission line are the redshifted values.
The blackbody luminosities are the unabsorbed values.
The interstellar absorption is modelled using \textsc{tbabs} with the abundances from \cite{wil00}.
One-sigma uncertainties are indicated for each parameter.
The uncertainties in the temperatures are calculated from simultaneous spectral and pulse profile fitting procedure described in Section 2.3.
 The uncertainties in the other spectral parameters are calculated using the {\tt XSPEC error} command.
}
\label{tbl:obs_params}
\end{table}

\section{Discussion}
\label{section:discussion}

The high resolution and statistics of the accumulated \xmm\ data on \pupa\ have allowed us to map the temperature profile of the surface of the NS in addition to constraining its viewing geometry.  They confirm the presence of two approximately antipodal spots, as found by
\citet{got10}, but also uncover the presence of an offset  $\Delta_{\rm antipodal}$ of at least  $1.\!^{\circ}1 \pm 0.\!^{\circ}2$ with respect to the (perfectly) antipodal configuration identified in the previous, lower-resolution study.
We calculated the expectation value of this offset $<\Delta_{\rm antipodal}> = 9.\!^{\circ}35 \pm 0.\!^{\circ}17$.

\subsection{Further Constraints on the Most Probable Geometries}
\label{subsection:fine_tuning}

We can draw further conclusions that certain emission geometries are very improbable if they require a ``fine tuning'' of the angle $\psi$ and the other model parameters.
Naturally, there should be no statistical correlation between the angle $\psi$ at which we are viewing \pupa, and its surface emission geometry.
For each degenerate solution, we can ask: ``What is the probability that a randomly selected angle $\psi$ would produce the observed, similar hot and warm spot fluxes?''
Referring to Table \ref{tbl:obs_params}, we see that the ratio of the \emph{observed} luminosities is:

\begin{equation}
\label{eq:equal_lum}
\frac{L_{\rm hot}}{L_{\rm warm}} = 1.06^{+0.05}_{-0.04}
\end{equation}

For some configurations of the hot and warm spots, there is a larger range of values of $\psi$ that satisfy this condition which we observe.
This is because in these geometries the hot and warm spot really do have similar intrinsic luminosities.
Alternatively, if the two spots really do have very different luminosities, then we would only measure similar observed fluxes if we are viewing \pupa\ from a more narrow range of viewing angles $\psi$.

For each degenerate solution, keeping all model parameters except $\psi$ fixed, we calculate the minimum and maximum values of $\psi$ that are consistent with Equation \ref{eq:equal_lum}.
The probability $P_{\rm config}$ of a given configuration is then proportional to the probability that a value of $\psi$ in this range will be drawn from a sinusoidal distribution:
\begin{equation}
P_{\rm config} \propto  \int^{\psi_{\rm max}}_{\psi_{\rm min}}  \sin\psi\ d\psi  
\end{equation}
Here $\psi$ is drawn from a sinusoidal distribution because of same the geometric argument described in Section  \ref{subsection:Most_probable_geometries}.
We find that there is a $\lesssim 1\%$ probability that $\xi_{h} < 1^{\circ}$ and a
$\lesssim 1\%$ probability that $\xi_{h} > 35^{\circ}$.
Finally, we find that the probability distribution function for the value of the hot spot colatitude peaks at $\xi_{h} \approx 6^{\circ}$. 

\subsection{Insensitivity of Results to Values of the NS Radius, Mass, and Distance}
\label{subsection:radius_distance_discusion}
We have modeled \pupa \ as a 1.4~$\msun$, 12~km radius neutron star at a distance of 1.3~kpc.
Some results of our modeling, like the values of $\beta_{w}$ and $\beta_{h}$, do depend on $M_{\rm NS}$, $R_{\rm NS}$ and $D_{1.3}$.
However, other model parameters are much less sensitive.
Most importantly, the angle $\delta_{\gamma}$ is completely  independent of $M_{\rm NS}$, $R_{\rm NS}$ and $D_{1.3}$, so we can be confident that have measured the correct minimum deviation from an antipodal geometry in any case.
Also, the curves in Figure \ref{fig:degeneracy}, tracing the possible hot spot configurations, are not particularly sensitive to the values of $M_{\rm NS}$, $R_{\rm NS}$ and $D_{1.3}$.
This is why, for example, the values of $(\delta_{\xi},\psi)$ for the two degenerate configurations where $\delta_{\xi}=0$, intersect the one-sigma confidence intervals of the perfectly antipodal solutions calculated in \cite{got10}, even though \cite{got10} assumed the NS was at the much larger distance of 2.2~kpc.

\subsection{Physical Explanation of the Surface Emission Geometry}
The most natural explanation for temperature anisotropies on the
surface of NSs involves the effect of crustal magnetic
fields, which leave their imprint on the surface temperature
distribution, and hence on the pulsed X-ray emission
\citep{gre83,gol92,pag95,hey98,pot01,lai01,hol02,gep04,cum04,
gep06,per06,zan06,pon09,gon10,per11, gla11,
pon11,vig13,per13,gep14,gou18,lan19,gou20,deg20,igo21,vig21}.  
The degree of temperature anisotropy is controlled by the ratio
between the thermal conductivity along and across the $B$-field lines.  In
the outer $\sim100$~m of a NS, the temperature gradient in the radial
direction is very high, and heat is efficiently transported if the $B$-field is in the radial
direction, thermally connecting the surface to the inner crust
and the core. On the other hand, crustal regions where the $B$-field
is nearly tangential are thermally insulated and not connected to the
hot core, hence remain cooler.

As the star ages, the magnetic field and the temperature evolve in a
coupled fashion. The Lorentz force in the induction equation causes
Hall drift of the field lines, while the Joule term is responsible
for Ohmic dissipation, which affects the temperature.  The
time-dependent temperature/magnetic field structure at a given age
depends on the macrophysics of the star, as well as on the
initial magnetic field configuration.  For a magnetic field which is
predominantly poloidal at birth, the temperature profile is symmetric
with respect to the equator, and the symmetry is maintained throughout
the evolution.  However, the presence of strong internal toroidal
components can radically change this topology. If the toroidal field
is dipolar (i.e. antisymmetric relative to the mirror-reflection about the equatorial plane) then the equatorial symmetry 
is broken due to the Hall drift
during the evolution (see, e.g., \citealt{hol02,vig13}), 
and it results in a complex field geometry with
asymmetric north and south hemispheres \citep{gep06,per13,gep14}.  In this case the
region with tangential field lines does not coincide with the equator,
resulting in asymmetric temperature profiles.  The degree of
anisotropy strongly depends on the initial toroidal field strength
because of its insulating effect in the crust.  Magnetothermal
simulations (e.g., \citealt{gep04,per13,gep14}) have shown that
temperature differences of more than a factor of 2 between the two
hemispheres can be produced in older, evolved objects.  Note
that magnetothermal simulations to date have been largely 2D (axial symmetry) for
computational reasons; hence the initial field configurations have
azimuthal symmetry, and so does the resulting temperature profile. 
Magnetothermal simulations in 3D have recently been carried out \citep{gou18,igo21} and shown that,  in models in which the dipolar field is misaligned with the toroidal field, the location of the hot spots is no longer aligned with the axis of the dipolar field, since the formation of the hot spots is very sensitive to the toroidal field.

Based on the discussion above, it is apparent that, at least at a qualitative
level, a surface temperature distribution of the kind inferred in \pupa\ is
plausible in NSs. However, there is an important caveat in the direct
application to \pupa: strong temperature
anisotropies require relatively large magnetic fields.
An investigation by \citet{gep14} into the formation of (magnetically-generated)
hot spots identified a number of important conditions for the formation of such
spots, including an initial surface dipolar field strength in the range $ 5\times 10^{12}\lesssim B_{\rm dip}\lesssim 5\times 10^{13}$~G
and an initial maximum crustal toroidal field $10^{15}\lesssim B_{\rm tor}\lesssim 6\times 10^{15}$~G.
While the latter is hidden in the crust and mainly revealed through its effects
on the surface temperature, the former is also independently measured from
the spin-down rate. In the case of \pupa, the field inferred from the spin down rate 
is $B_s = 2.9 \times 10^{10}$~G, which is in apparent tension with the above results taken at
face value (unless there has been a considerable field decay).  Possible ways to ameliorate this tension may be for
\pupa\ to  have a poloidal field with stronger-than-dipolar multipoles and/or the presence of material (such as a 
fallback disk left over from the supernova explosion) which provides additional
torque to the star, and may hence affect the field measurement from spin down alone
(see, e.g., \citealt{yan12}).

Alternatively, the temperature anisotropy observed in \pupa\ (as well
as its spin-down properties) could be interpreted within the context of a
``buried'' field scenario \citep{gep99,ho11,sha12,vig12}.  The NS does
not need to be born with a very weak magnetic field, like in the
anti-magnetar scenario, but it could have a typical field which is
buried by an episode of hypercritical accretion following the
supernova explosion. This leads to an external magnetic field
(responsible for the spin-down of the star) much weaker than the
internal ``hidden'' $B$ field.  This model has the distinctive feature
of producing hot polar caps since the buried $B$ field keeps the
equatorial surface regions insulated from the hot core \citep{vig12}.
The additional advantage of this scenario is that, unlike in the standard
magnetothermal evolution where hot spots form on a timescales of at least $\sim 10$~kyr,
the hot regions form on a much shorter timescale, as little as $\sim 1$~kyr,
hence more compatible with the young age of $\sim4.6$~kyr inferred for \pupa.

\subsection{Comparison with Other CCOs}
It is interesting to compare \pupa\ to \kes, the CCO in the Kes 79 supernova remnant.
\cite{bog14} also used a general relativistic modeling technique to study the pulsations from \kes\ that are, unlike \pupa, strongly pulsed and significantly broader than a pure sinusoid.
This indicates that either the emission is strongly beamed, or the heated surface region has a very elongated, non-circular shape.

The only other CCO with detected X-ray pulsations is \tos, which has a small pulsed fraction of $\sim10\%$.
There is no evidence of an energy dependent phase reversal in its pulse profiles.
Evidently the two-temperature spectral components are approximately co-located on the stellar surface.
Because \tos\ only has one surface thermal region producing its observed pulse profile, we can conclude that it likely has a relatively small magnetic inclination angle.

A small (or undetectable) pulsation amplitude is common to all the CCOs (with the exception of \kes\ in Kes 79), and hence
something potentially interesting as a diagnostic of their nature \citep{got13}.
In the case of Puppis~A, we have calculated that it is likely the result of a small inclination angle between
the emitting regions and the rotation axis. In the buried field scenario,
the re-emergence of the field occurs on a timescale $\sim 10^3-10^5$~kyr.
Simulations by \citet{vig12} for a representative case with an internal toroidal field
of $10^{15}$~G, and a partially re-emerged dipolar field of $10^{10}$~G show that,
at an age of a few kyr and for a crustal-confined magnetic field model, the NS
displays hot polar caps. Hence,	in this	scenario, the hot regions correspond
to the axis of the dipolar field. The small inclination	axis between the dipolar and
rotation axis can then be interpreted
within the evolutionary	model of \citet{dal17}, which couples interior NS viscosity
and magnetic field evolution to	predict	the expected range of inclination angles of young neutron stars.
At the 112~ms period of \pupa, a small inclination angle is predicted for NSs with magnetically induced
ellipticities $\epsilon_B\gtrsim$~a few $\times 10^{-7}$, with the specific value dependent
on the mass and	radius of the NS. Hence, a large internal field in the hidden field scenario
can explain both the observed anisotropic temperature distribution and the small observed inclination angle.

\subsection{Limitations, and Directions for Future Work}
This study used blackbodies to model the emission from the surface of \pupa.
This yielded a very good reproduction of the observed X-ray spectrum and pulsations.
However, it is possible that a NS atmosphere model could better reproduce some of the fine details that may be missing from a pure blackbody model.
For example, at some energies our model over or under-predicts the observed pulse amplitudes.
Anisotropic emission from a NS atmosphere could be produce these small changes in pulse amplitude relative to the isotropic case.
With the current data, it is not clear if the differences between the data and isotropic emission is just statistical noise.
If a NS atmosphere model is appropriate, then it would imply that the sizes of the hot and warm spots are systematically larger than the blackbody sizes computed in this study.
This is because a blackbody is the most efficient emitter at a given effective temperature.
Finally, we remark that future X-ray polarization measurements may be able to both further constrain the viewing geometry, and also probe the magnetic field near the two surface emitting regions of \pupa.

\section{Summary}
CCOs are a common class of young NSs, suspected to have internal magnetic fields much stronger than their external (dipolar) ones. Here we have reported the results of our analysis of $471$~ks of \xmm\ observations of a special member of this class, the CCO in the Puppis~A supernova remnant. We used the wealth of high-resolution \xmm\ data to perform a detailed analysis of the energy-dependent pulse profile in sixteen energy bands.
Our modeling, which accounts for the general relativistic effects of light bending and gravitational redshift, precisely measured the (redshifted) temperatures  on the star surface, in addition to constraining the viewing/emission geometry.

We uncovered an asymmetric temperature distribution:  a warm surface region with a temperature $kT_{\rm warm} = (1+z) \times  0.222_{-0.019}^{+0.018}$ keV, and a hotter region with $kT_{\rm hot} = (1+z) \times  0.411\pm0.011$  keV, offset from the antipodal position by at least $\Delta_{\rm antipodal} = 1.\!^{\circ}1 \pm 0.\!^{\circ}2$.
Averaging over the most probable emission geometries, we calculated the expectation value of this offset $<\Delta_{\rm antipodal}> = 9.\!^{\circ}35 \pm 0.\!^{\circ}17$.
We presented the full range of degenerate emission geometries, and calculated that the most likely geometries feature a small hot spot close to the rotational pole.
We then discussed the production of such a small hot spot
within the context of the anti-magnetar scenario or, alternatively, as arising within a buried field scenario.  
In either case, the anisotropic heat conduction on the NS surface seems to require crustal magnetic fields that are stronger than, and misaligned with the external spin-down measured field.

\begin{acknowledgements}
JAA and EVG acknowledge support from NASA grants 80NSSC20K0046 and 80NSSC20K0717, and from SAO grant G09-20055X.
RP acknowledges support from the NSF through the award NSF-AST 2006839.
This research made use
of data and software provided by the High Energy Astrophysics Science
Archive Research Center (HEASARC), which is a service of the
Astrophysics Science Division at NASA/GSFC and the High Energy
Astrophysics Division of the Smithsonian Astrophysical Observatory. We
also acknowledge use of the NASA Astrophysics Data Service (ADS).
\end{acknowledgements}

\facility{XMM}

\software{{\tt astropy} \citep{ast13, ast18}, {\tt matplotlib} \citep{hun07}, {\tt xspec} \citep{arn96}, {\tt numpy} \citep{har20}, {\tt scipy} \citep{vir20}}

\appendix

 \section{\xmm \ pn Detector Background Subtraction}

\begin{figure*}[h]
\centering
\includegraphics[width=1.0\textwidth]{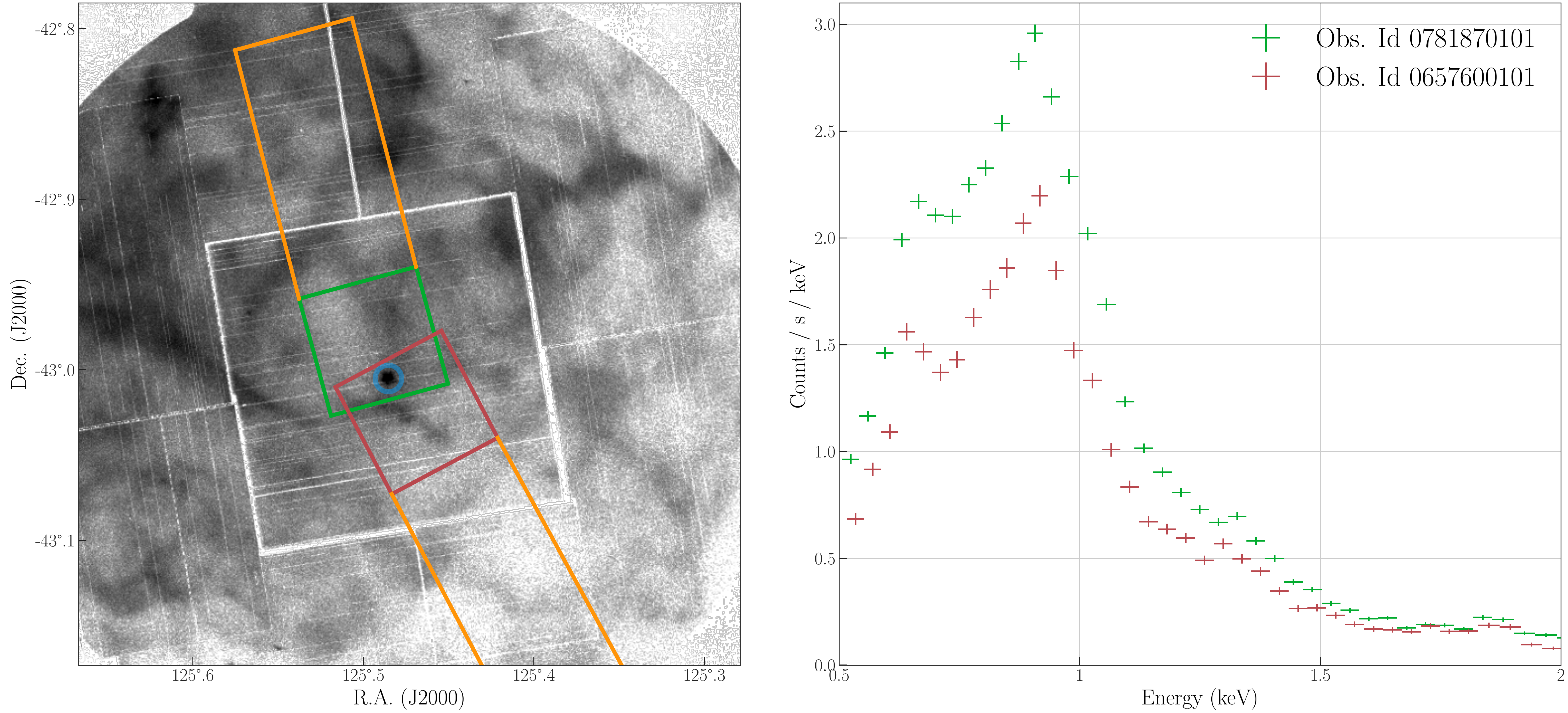}
\caption{Left: Orientation of the \xmm\ pn detector when operating in SW mode during observations 0781870101 and 0657600101. 
The red and green squares outline the edges of the SW region, and the orange rectangles mark the rest of the CCD.
Events detected in the orange region of the CCD can contribute to the background in the SW regions.
We extracted the background photons from the same blue annular region (inner/outer radii of $32^{\prime\prime}$/$45^{\prime\prime}$, centered on the CCO) in both observations.
The X-ray background image shown in this figure was created by combining the two MOS detector images from the 0781870101 observation.
Right: The background spectra from both observations. 
The background rate is higher during observation 0781870101 because the CCD is oriented such that there is more contamination from where the orange region intersects the brighter section of the SNR.
}
\label{fig:appendix_1}
\end{figure*}

When the \xmm\ pn detector is operated in small window (SW) mode, the entire CCD, not just the small window, is illuminated.
In this configuration there are out-of-time events  from the CCD region outside of the small window.
Since the CCD region outside the small window sometimes detects photons from the Puppis A supernova remnant, the background count rate is roll-angle dependent.
The roll angle is a function of the position of \xmm\ relative to the Sun, and background count rates are higher during observations performed in the months of April and May.
Figure~\ref{fig:appendix_1} shows the location of the pn detector during two representative observations with high and low background rates.

We tested for the possibility of imperfect background subtraction by experimenting with different background scaling factors.
We implemented this background scaling using the HEASoft tool {\tt grppha}.
The background scaling factor was stepped through values of 0.80 to 1.20 in steps of 0.01.
All datasets were combined into one spectrum with the HEASoft tool {\tt addascaspec}.
We also divided the datasets into two groups with high and low roll angles, and combined the spectra in these two separate groups.
These three combined spectra, as well as each individual spectrum, were fitted with simple two blackbody models, i.e. without an emission or absorption line feature, and the $\chi^2$ values were calculated for each value of the background scaling factor.
We found that, in all cases, the fit to the spectrum is worse with the pure two blackbody model than when the model includes an emission line feature.
If the emission feature was due to contamination from the SNR, our variable background scaling procedure would have allowed us to achieve a better fit to a pure blackbody spectrum for some value of the background scaling factor.
Because the inclusion of the line results in better fits to the source spectrum in all cases, we conclude the emission line feature is intrinsic to the source spectrum.

Next, simultaneously fitting a two blackbody plus Gaussian emission line model to all 19 observations, we looked for variability in the emission feature.
The column density of interstellar absorbing material and the Gaussian line width were held constant across all observations, and the other parameters were allowed to vary independently.
The photo-electric absorption by the interstellar medium was modeled with the Tuebingen-Boulder ISM absorption model, implemented as {\tt tbabs} in {\tt XSPEC}.
For each observation, the residuals of the fit of the models to the data were small and showed no need for any additional model components.
The central energy of the Gaussian line is $\approx0.75$~keV.
We find that the emission-line central energy, width, and normalization are, to within statistical uncertainties, invariant between observations.
In particular we found no significant difference in the source spectra of the observations with different background rates.
In summary, the inclusion of the emission line feature results in optimal fits to each dataset, regardless of background rates, and it can be modeled as a time-invariant to within statistical uncertainties of the data.

\section{Derivation of Equation (9), Intersection of two spherical caps}

Given two spherical caps with angular radii r$_{1}$ and $r_{2}$, with centers separated by an angle $\theta$ on a unit sphere, the area of intersection can be calculated with elementary spherical trigonometry (see, e.g., Chapter 1 of \citealt{sma77}).  
A schematic planar representation is shown in Figure~\ref{fig:appendix_2}.
The centers of the spherical caps are labeled A and B. 
Cap A has a spherical radius $r_{1}$ and Cap B has a spherical radius $r_{2}$.
The two circles outline the boundaries of the spherical caps and the straight lines correspond to arcs of great circles on the unit sphere.
As shown in the bottom of the figure, the area of the intersection of the two spherical caps is given by the sum of the areas of the two spherical sectors minus the areas of the two identical spherical triangles.
The areas of the two spherical sectors, denoted $\sector_{A}$ and $\sector_{B}$, are given by:

\begin{figure*}[h]
\centering
\includegraphics[width=0.9\textwidth]{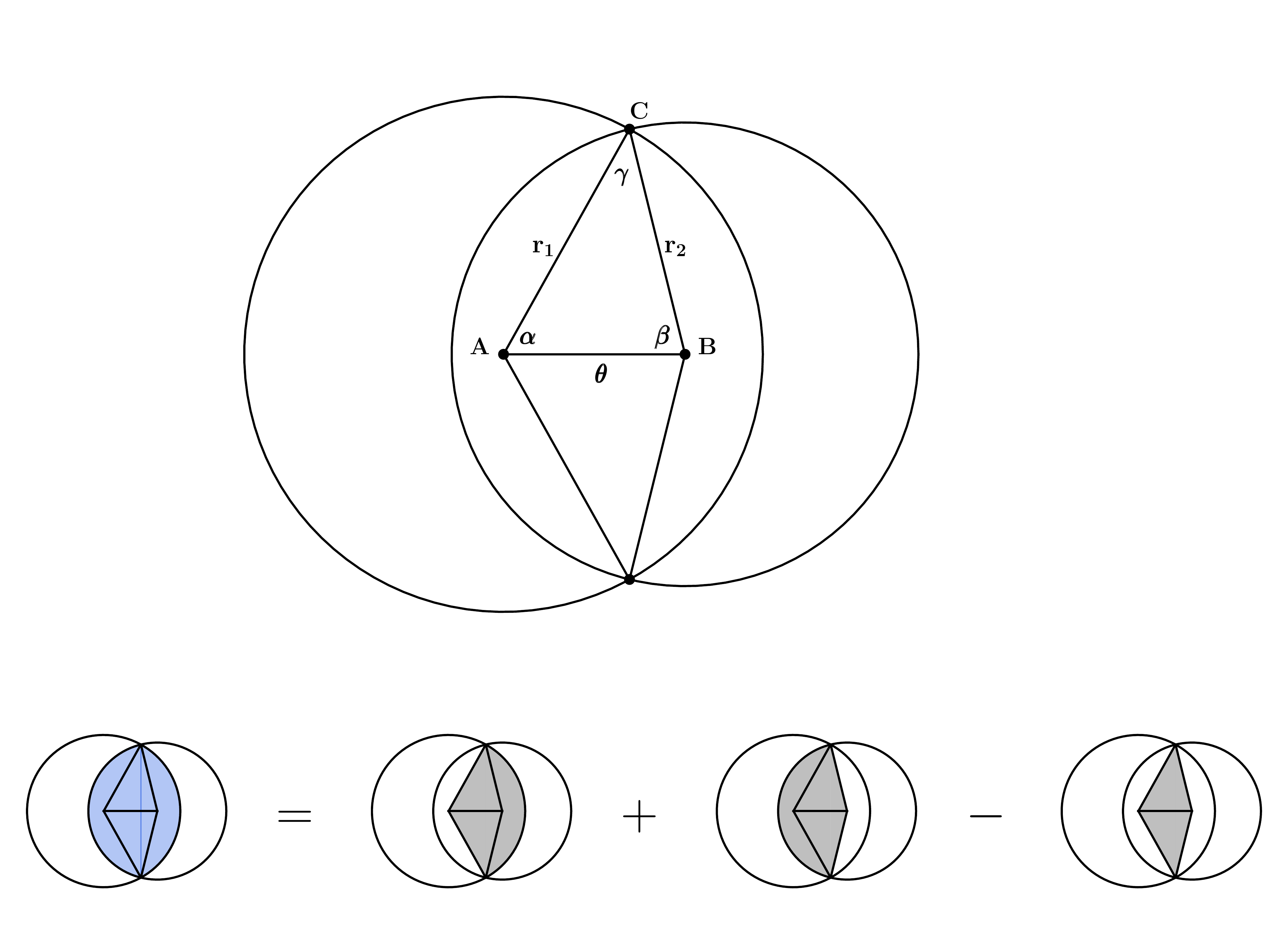}
\caption{Top: Schematic planar representation of the intersection of two spherical caps. One of the caps represents the region of the NS that is invisible to an observer, and its size is a function of the mass and radius of the NS. The other cap represents the thermally emitting region on the NS surface. Bottom: The area of region of intersection is computed as a combination of spherical sectors and spherical triangles as shown.} 
\label{fig:appendix_2}
\end{figure*}

\begin{equation}
    \sector_{A} = 2 \alpha (1 - \mathrm{cos}\ r_{1})
\end{equation}

and 

\begin{equation}
    \sector_{B} = 2 \beta (1 - \mathrm{cos}\ r_{2})
\end{equation}

The arcs connecting the centers of the spherical caps and the intersection of the spherical caps form two identical spherical triangles, with sides of length $r_1$, $r_2$, and $\theta$.
The area of a spherical triangle on a unit sphere is equal to the sum of its angles minus $\pi$.
So the area $A_{\triangle}$ of each of the two identical spherical triangles is equal to 

\begin{equation}
    A_{\triangle} = \angle CAB +  \angle ABC + \angle BCA  - \pi = \alpha + \beta + \gamma - \pi 
\end{equation}

We use the spherical law of cosines to write the angles $\alpha$, $\beta$, and $\gamma$ in terms of the arc lengths $r_1$, $r_2$, and $\theta$:

\begin{equation}
    \alpha = \cos^{-1}\bigg(\frac{\mathrm{cos}\ r_2} { \mathrm{sin}\ \theta\ \mathrm{sin}\ r_1} -
  \mathrm{cot}\ \theta\ \mathrm{cot}\ r_1 \bigg) 
\end{equation}

\begin{equation}
    \beta = \cos^{-1}\bigg(\frac{\mathrm{cos}\ r_1} { \mathrm{sin}\ \theta\ \mathrm{sin}\ r_2} -
  \mathrm{cot}\ \theta\ \mathrm{cot}\ r_2 \bigg) 
\end{equation}

\begin{equation}
    \gamma = \cos^{-1}\bigg(\frac{\mathrm{cos}\ \theta} { \mathrm{sin}\ r_1\ \mathrm{sin}\ r_2} -
  \mathrm{cot}\ r_1\ \mathrm{cot}\ r_2 \bigg) 
\end{equation}

We can now compute the area of intersection of the two spherical caps:

\begin{equation}
I(\theta, r_1, r_2) = \sector_{A} + \sector_{B} - 2 A_{\triangle} = 2\pi - 2 \gamma - 2\alpha\ \mathrm{cos}\ r_1 - 2\beta\ \mathrm{cos}\ r_2
\end{equation}

Inserting the expressions for $\alpha$, $\beta$, and $\gamma$ from equations B4 through B6 gives the equivalent of Equation (9) in the text:

\begin{multline} 
I(\theta, r_1, r_2) = 2 \pi 
-2 \cos^{-1}\bigg(\frac{\mathrm{cos}\ \theta}
  { \mathrm{sin}\ r_1\ \mathrm{sin}\ r_2} -
  \mathrm{cot}\ r_1\ \mathrm{cot}\ r_2 \bigg) 
\ - \ 2 \cos^{-1}\bigg(\frac{\mathrm{cos}\ r_2}
  { \mathrm{sin}\ \theta\ \mathrm{sin}\ r_1} -
  \mathrm{cot}\ \theta\ \mathrm{cot}\ r_1 \bigg) \mathrm{cos}\ r_1  \\
 \  - \ \ 2 \cos^{-1}\bigg(\frac{\mathrm{cos}\ r_1}
  { \mathrm{sin}\ \theta\ \mathrm{sin}\ r_2} -
  \mathrm{cot}\ \theta\ \mathrm{cot}\ r_2 \bigg)
  \mathrm{cos}\ r_2
\label{eq:cap_intersection_appendix}
\end{multline}

\end{document}